\documentclass{elsart}
\usepackage{epsfig,amsmath,amssymb}
\newcommand{\EEE}{\mathbf{E}}
\newcommand{\DDD}{\mathbf{D}}
\newcommand{\HHH}{\mathbf{H}}
\newcommand{\rrr}{\mathbf{r}}
\newcommand{\nablabf}{\mathbf{\nabla}}
\newcommand{\varepsilonbf}{\boldsymbol{\varepsilon}}

\begin{document}

\begin{frontmatter}

\title{Non-perturbative approach to high-index-contrast variations in electromagnetic systems}

\author{Lars Rindorf$^1$ and Niels Asger Mortensen$^2$}

\address{$^1$COM$\bullet$DTU,
Technical University of Denmark,\\ Building 345 West, \O rsteds
Plads, DK-2800 Kongens Lyngby, Denmark,\\
$^2$MIC--Department of Micro and Nanotechnology, NanoDTU,\\
Technical University of Denmark,\\ Building 345 East, \O rsteds
Plads, DK-2800 Kongens Lyngby, Denmark}

\begin{abstract}
We present a method that formally calculates \emph{exact}
frequency shifts of an electromagnetic field for arbitrary changes
in the refractive index. The possible refractive index changes
include both anisotropic changes and boundary shifts. Degenerate
eigenmode frequencies pose no problems in the presented method.
The approach relies on operator algebra to derive an equation for
the frequency shifts, which eventually turn out in a simple and
physically sound form. Numerically the equations are well-behaved,
easy implementable, and can be solved very fast. Like in
perturbation theory a reference system is first considered, which
then subsequently is used to solve another related, but different
system. For our method precision is only limited by the reference
system basis functions and the error induced in frequency is of
second order for first-order basis set error. As an example we
apply our method to the problem of variations in the air-hole
diameter in a photonic crystal fiber.
\end{abstract}

\end{frontmatter}

%------------------------------------------------------------------
\section{Introduction}
%------------------------------------------------------------------

Scale invariance of Maxwell's equations provides unique and simple
relations between the electromagnetic properties of dielectric
structures which only differ by a simple spatially uniform scaling
of the dielectric function~\cite{joannopoulos}. The frequency
shifts of electromagnetic modes due to spatially non-uniform
low-index-contrast changes have typically successfully been
addressed with perturbation theory (see e.g.
Ref.~\cite{marcuse1991}) being quite similar to the one used
widespread in quantum mechanics (see e.g.
Ref.~\cite{ballentine1998}). However, the development of fields
such as photonic band-gap materials~\cite{joannopoulos,knight2003}
and surface-plasmon subwavelength
optics~\cite{barnes2003,pendry2004b} calls for an understanding of
high-index-contrast dielectric variations beyond standard
perturbation theory. There has recently been an effort toward a
perturbative description of high-index-contrast problems with
shifting material boundaries
\cite{johnson2002,skorobogatiy2002a,skorobogatiy2003a,skorobogatiy2004a}.
In Ref.~\cite{johnson2002} the problem is addressed by approaching
the strong dielectric contrast as a small perturbation in the
material boundary. This different approach to standard
perturbation theory uses additional ideas from quantum mechanics,
such as multiplying the perturbing dielectric function by a
coupling strength variable. The related problem of scattering of
electromagnetic waves by high-index-contrast dielectric structures
has been addressed by essentially non-perturbative approaches
relying on iterative solutions of the Dyson equation and the
Lippmann--Schwinger equation for the electromagnetic Green's
function and the electromagnetic fields,
respectively~\cite{greffet1988,martin1995,Rahachou2005a}.

Here, we describe a non-perturbative approach to electromagnetic
eigenmodes that formally calculates \emph{exact} frequency shifts
of an electromagnetic field for arbitrary changes in the
refractive index. As an example we apply our theory to photonic
crystal fibers~\cite{knight1996} and show how the approach can be
used to study the effect of variations in the air-hole diameter.

The paper is organized as follows: In Sec.~\ref{sec:formalism} we
present the formalism and in Secs.~\ref{sec:vare} and
\ref{sec:vare} we present the variational method in electric and
magnetic field. In Sec.~\ref{sec:example} we apply our theory to
photonic crystal fibers. Finally, in Sec.~\ref{sec:conclusion}
discussions and conclusions are given.

%------------------------------------------------------------------
\section{General formalism}\label{sec:formalism}
%------------------------------------------------------------------
%------------------------------------------------------------------
\subsection{The temporal harmonic problem}
%------------------------------------------------------------------
In the following we consider temporal harmonic solutions to
Maxwell's equations. The electrical field and magnetic field may
then be written as
\begin{eqnarray}
\EEE(\rrr,t) = \EEE_\omega(\rrr)e^{-i\omega t},\\
\HHH(\rrr,t) = \HHH_\omega(\rrr)e^{-i\omega t},
\end{eqnarray}
with the subscript $\omega$ indicating the frequency dependence.
The electrical field is a solution to a vectorial wave equation,
which has the form of a generalized eigenvalue
problem~\cite{joannopoulos,sakoda},
\begin{eqnarray}
 \nablabf \times
\nablabf \times \EEE_\omega(\rrr) =
\epsilon_0\boldsymbol{\varepsilon}(\rrr)\Big(\frac{\omega}{c}\Big)^2\EEE_\omega(\rrr), \label{eigenvalueE}
\end{eqnarray}
where $\boldsymbol{\varepsilon}$ is the $3\times3$ dielectric
tensor. The corresponding eigenvalue problem in the magnetic field
has the form~\cite{joannopoulos,sakoda}
\begin{eqnarray}
 \nablabf \times
\epsilon_0^{-1}\boldsymbol{\varepsilon}^{-1}(\rrr) \nablabf \times
\HHH_\omega(\rrr) = \Big(\frac{\omega}{c}\Big)^2\HHH_\omega(\rrr).
\label{eigenvalueH}
\end{eqnarray}
We consider lossless dielectrics so that
$\boldsymbol{\varepsilon}$ is real and symmetric positive
semi-definite, i.e. it has non-negative eigenvalues. In the
following we will for simplicity put $c=\epsilon_0=1$ and omit the
subscripts on $\EEE_\omega$ and $\HHH_\omega$.

%------------------------------------------------------------------
\subsection{Operator formalism}
%------------------------------------------------------------------
In analogy with quantum mechanics we will in this work take
advantage of a function space known as a Hilbert space, which
consists of a set of basis functions and an inner product. In the
following we use the Dirac \emph{bra-ket} notation and define the
inner product as
\begin{eqnarray}
\big< A_m \big|  A_n \big> = \int d\rrr\,
\mathbf{A}_m^\dagger(\rrr)\cdot \mathbf{A}_n(\rrr)\label{norm}
\end{eqnarray}
where $\dagger$ is the Hermitian conjugate (transpose and complex
conjugate) of the field vector. If the basis of the Hilbert space
is complete then any function in the Hilbert space can be written
as a linear combination of basis functions, thus we can rewrite
any operator $\hat{O}$ as
\begin{eqnarray}
\hat{O} = \sum_{mn} O_{mn} \big|A_m\big>\big<A_n\big|,
\end{eqnarray}
where $\hat{O}$ has matrix elements
\begin{eqnarray}\label{eq:Omn}
O_{mn} \equiv  \big< A_m \big|\hat{O} \big| A_n
\big> = \int d\rrr\, \mathbf{A}_m^\dagger(\rrr)
\mathbf{O}(\rrr) \mathbf{A}_n(\rrr).
\end{eqnarray}
Typically, the basis is not complete, but in many practical
applications this is not imperative. Note,  that $\nablabf \times
$ and $\hat{\varepsilon}$ are no longer a differential operator
and a tensor, respectively, but linear operators in the Hilbert
space.

%-------------------------------------------------------------
\subsection{The reference system}
%-------------------------------------------------------------
We will in the following consider a system that is similar to the
systems we are trying to solve. We refer to it as the `reference'
system, and it is described by a dielectric system with dielectric
function $\varepsilon_0$, and its modes all have frequency
$\omega_0$, but possibly with different wave vectors $\boldsymbol
k$ . The corresponding operators will be assigned a subscript `0'.
The eigenmodes are determined by either Eq.~(\ref{eigenvalueE}),
or Eq.~(\ref{eigenvalueH}).

%-------------------------------------------------------------
\section{Variational approach in $\EEE$}\label{sec:vare}
%-------------------------------------------------------------
From Eq.~(\ref{eigenvalueE}) it follows that the frequency of the
field $|E\big>$ can be determined by
\begin{equation}
\omega^2 = \frac{\big<E\big|\nablabf \times \nablabf \times \big|
E\big>}{\big<E\big|\hat{\varepsilon}\big|E\big>}
\end{equation}
and since both $\nablabf \times \nablabf \times$ and $\hat{\varepsilon}$
are Hermitian this forms the typical starting point for
variational calculations. Here, we will show a different route
which is beneficial for electromagnetic modes with the same
frequency, but possibly with different wave vectors $\boldsymbol
k$.

We may choose the eigenmode basis of the reference system,
$\big\{E_n\big\}_{n=1}^N$, where eigenmodes are orthogonal with
the norm
\begin{eqnarray}
\big< E_m\big| \hat{\varepsilon}_0\big| E_n \big> = \delta_{mn}.
\end{eqnarray}
Here, $\hat{\varepsilon}_0$ is the dielectric function operator and
$\delta$ is the Kronecker delta function.

In the spirit of perturbation theory we now solve a different but
assumed similar system using the reference eigenmodes and
frequencies. In quantum mechanics such a system is for historical
reasons often referred to as the \emph{interacting system} though
the approach of course also can be applied to one-particle
problems. Here we will decline from using the term
\emph{interacting system} and rather refer to it as the \emph{new
system}, since we seek to avoid linking our method to approximate
perturbation theory. For the new system we denote the
eigenfrequencies by $\omega$ and the new system also obeys the
eigenvalue equation, Eq.~(\ref{eigenvalueE}), and is described by
the dielectric function, $\varepsilonbf(\rrr)$. It is important to
note that the frequencies of the new system do not need to be
identical. In the new basis (eigenstate of reference system $n$)
we define
\begin{eqnarray}\label{eq:T_newsystem_crude}
\nablabf \times \nablabf \times \big| E_n\big> =
\hat{\varepsilon}\hat{\Omega}^2\big| E_n\big>. \label{eigenNew}
\end{eqnarray}
The double curl operation acting on a reference system eigenmode
can be replaced by $\hat{\varepsilon}_0\hat{\omega}_0^2$ and the
equation rewritten as
\begin{eqnarray}\label{eq:T_newsystem}
\hat{\varepsilon}_0\hat{\omega}_0^2| E_n\big> =
\hat{\varepsilon}\hat{\Omega}^2\big| E_n\big>  \Leftrightarrow
\hat{\varepsilon}_0\hat{\Omega}^2\big| E_n\big> =
\hat{\varepsilon}_0\hat{\varepsilon}^{-1}\hat{\varepsilon}_0
\omega_{0}^2\big| E_n\big>,
\end{eqnarray}
where we have replaced the operator $\hat\omega_0^2$ by the
corresponding eigenvalue $\omega_{0}^2$. Next, we define
\begin{eqnarray}
\hat{T} \equiv
\hat{\varepsilon}_0\hat{\varepsilon}^{-1}\hat{\varepsilon}_0 =
\hat{\varepsilon}_0\frac{\hat{\Omega}^2}{ \omega_0^2}, \label{T}
\end{eqnarray}
so that our problem, Eq.~(\ref{eigenvalueE}), can be rewritten as
\begin{eqnarray}\label{eq:newsystem}
\omega_0^2\hat{T}\big|E\big>=\hat{\varepsilon}_0 \omega^2 |E\big>.
\end{eqnarray}
Obviously, $\omega_0^2\hat{T}$ is a Hermitian operator on our
Hilbert space since
$\big(\hat{\varepsilon}_0\hat{\varepsilon}^{-1}\hat{\varepsilon}_0\big)^\dagger
= \hat{\varepsilon}_0\hat{\varepsilon}^{-1}\hat{\varepsilon}_0$.
Similarly, $\hat{\varepsilon}_0 \omega^2$ is also Hermitian since
$\hat{\varepsilon}_0$ is Hermitian.

%------------------------------------------------------------------
\subsection{The variational problem}
%------------------------------------------------------------------
We know expand the \emph{new system} eigenmode(s) in the basis of
reference system eigenmodes,
\begin{eqnarray}
\big| E \big>  = \sum_{n} c_n \big| E_n \big>,
\end{eqnarray}
and since we have Hermitian operators on both the left-hand and
right-hand sides of Eq.~(\ref{eq:newsystem}) we get the
variational problem
\begin{eqnarray}
 \omega^2\big[ E \big]
&=& \omega_0^2 \min_{\left| E \right>} \frac{\big<E \big|\hat{T}
\big|
E \big>}{\big<E \big|\hat{\varepsilon}_0\big| E \big>}\nonumber\\
& =& \omega_0^2 \min_{\{c_n\}} \sum_{nm} c_m^* c_n \big<E_m
\big|\hat{\varepsilon}_0\hat{\varepsilon}^{-1}\hat{\varepsilon}_0\big|
E_n \big>, \label{varE}
\end{eqnarray}
where $c_n$'s are normalized so that $\sum_n |c_n|^2=1$. It is of
course interesting to know how much error in frequency an error in
the state vector induces. For the numerical implementation this is
particular important since here the basis set is rarely complete.
Due to the variational principle, a first order error in the
eigenmode $\EEE (\rrr )$ will induce only a second error in
frequency $\omega$.

%------------------------------------------------------------------
\subsection{Finding ${\omega}$}
%------------------------------------------------------------------
We now want to find the eigenfrequencies of the new system, which
we denote $\omega_m$, $m=1,2,\dots ,N$. First we find the matrix
elements of the operator Eq.~(\ref{eq:T_newsystem}) as
\begin{eqnarray}
T_{mn}&=& \big<E_m
\big|\hat{\varepsilon}_0\hat{\varepsilon}^{-1}\hat{\varepsilon}_0\big|
E_n \big>
\nonumber\\
&=& \int d\rrr \, \EEE_m^\dagger(\rrr) \varepsilonbf_0(\rrr)
\varepsilonbf^{-1}(\rrr) \varepsilonbf_0(\rrr)
 \EEE_n(\rrr)
\nonumber\\
&=& \int d\rrr \, \DDD_m^\dagger(\rrr)
\varepsilonbf^{-1} (\rrr)
 \DDD_n(\rrr),
\end{eqnarray}
which may be expressed both in terms of $\EEE$ and $\DDD$ fields.
The variational problem Eq.~(\ref{varE}) can be solved by finding
the eigenvalues and eigenvectors of the matrix $\omega^2\hat{T}$.
Formally we thus have
\begin{eqnarray}
\hat\Omega^2 &=& \hat{C}^\dagger \hat{\omega}^2 \hat{C}
\label{Omega}
\end{eqnarray}
and the `new' eigenfrequencies are given by
\begin{eqnarray}
\hat{\omega} & =& \sqrt{\hat{C} \hat{\Omega}^2 \hat{C}^\dagger}=
\hat{C}  \hat{\Omega} \hat{C}^{\dagger}, \label{storomega}
\end{eqnarray}
where $\hat{\omega}$ is a diagonal eigenfrequency matrix and
$\hat{C}$ is a unitary transformation matrix,
$\hat{C}\hat{C}^\dagger=1$. The `new' system eigenstate vectors now
become
\begin{eqnarray}
\big| \tilde{E}_n \big> &=& \sum ^N _{m=1} C_{mn}\big|E_m\big>.
\end{eqnarray}
Being interested in the frequency shifts rather than the new
frequencies, one subtracts the reference frequencies from the new
frequencies defined by the eigenvalues of the operator given by
Eq.~(\ref{Omega}). This gives
\begin{eqnarray}
\Delta \omega_m &=& \omega_{m}-\omega_0.
\label{deltaomega}
\end{eqnarray}
It should be emphasized that in general mode $m$ in the reference
system may not be similar to the mode $m$ in the new system. This
is due to the fact that changes in the refractive indices may
favor one mode more than another, causing different effective
refractive indices of the modes to converge and then diverge again
with the difference now having opposite sign of before.

%------------------------------------------------------------------
\subsection{Relation to  first-order perturbation theory}
\label{sec:1pt}
%------------------------------------------------------------------
Consider a small variation in the dielectric function,
$\delta\hat{\varepsilon}=\hat{\varepsilon}-\hat{\varepsilon}_0$.
This variation is assumed small enough to be neglected beyond
first order. If the operators $\delta\hat{\varepsilon}$ and
$\hat{\varepsilon}^{-1}$ commute,
$[\delta\hat{\varepsilon},\hat{\varepsilon}^{-1}]=0$, (a
sufficient, but not necessary, condition for this commutator to be
zero is the two dielectric tensors being isotropic) the elements
of the square frequency operator $\omega_0^2\hat{T}$ are
\begin{eqnarray}
\omega_0^2T _{mn}   &=& \omega_0^2 \delta_{mn} + \omega_0^2
\big<E_m\big| \Big\{
\hat{\varepsilon}_0\big(\hat{\varepsilon}_0+\delta\hat{\varepsilon}\big)^{-1}\hat{\varepsilon}_0
- \hat{\varepsilon}_0 \Big\} \big|E_n \big>\nonumber\\
& =& \omega_0^2\delta_{mn} -\omega_0^2\big<E_m\big|
\delta\hat{\varepsilon}\big|E_n \big> +
\mathcal{O}(\delta\varepsilon^2). \label{1ptexpansion}
\end{eqnarray}
Expanding $\hat{\Omega} = \omega_0\sqrt{\hat{T}^2}$ in
$\delta\hat{\varepsilon}=0$ gives
\begin{eqnarray}
\Omega _{mn} = \omega_0 \delta_{mn} -
\frac{\omega_0}{2}\big<E_m\big| \delta\hat{\varepsilon}\big|E_n
\big> + \mathcal{O}(\delta\varepsilon^2) \label{1pt}
\end{eqnarray}
where the new frequencies are the eigenvalues of $\hat{\Omega}$. This is
the well-known result of first-order (degenerate) electromagnetic
perturbation theory (see e.g.
Refs.~\cite{marcuse1991,johnson2002}).

%-------------------------------------------------------------
\section{Variational approach in $\HHH$}\label{sec:varh}
%-------------------------------------------------------------
The $\HHH$-fields satisfy eigenvalue equation
Eq.~(\ref{eigenvalueH}) and all modes of the reference system
again share the same frequency $\omega_0$, and satisfy the
orthonormality condition $\big< H_m \big|H_n \big> = \delta_{mn}$.
Next, we consider a presumably similar system with a different
dielectric function, and expand the modes of the `new' system in
the basis functions of the reference system:
\begin{eqnarray}
\hat{\Theta} \big|H_n \big>  \equiv
\nabla\times  \hat{\varepsilon}^{-1} \nabla\times \big|H_n \big>  =
\nabla\times \hat{f} \hat{\varepsilon}_0^{-1} \nabla\times \big|H_n \big>  =
\hat{\Omega}^2 \big|H_n \big> ,
\label{Theta}
\end{eqnarray}
where we have introduced the operator $\hat{f} \equiv
\hat{\varepsilon}^{-1}\hat{\varepsilon}_0$. Taking the dielectric
functions to be anisotropic $\mathbf{f}(\rrr)=
\mathbf{1}_{3\times3}\varepsilon^{-1}(\rrr)\varepsilon_0(\rrr)$
and using vector analysis we may rewrite the expression as
\begin{eqnarray}
\nabla\times \hat{f} \hat{\varepsilon}_0^{-1} \nabla\times
\big|H_n \big>  &=& \nabla\hat{f} \times \hat{\varepsilon}_0^{-1}
\nabla\times \big|H_n \big> + \hat{f}\nabla\times
\hat{\varepsilon}_0 ^{-1}\nabla\times \big|H_n \big>\nonumber\\
& =& -i\omega \nabla\hat{f} \times \big|E_n \big> +
\hat{f}\omega^2 \big|H_n \big> ,
\end{eqnarray}
and the eigenvalue equation (\ref{Theta}) thus reduces to the
operator equation where $\hat{\Theta}$ has matrix elements
\begin{eqnarray}
\Theta_{mn} =
-i\omega_0 \big< H_m \big|\nabla\hat{f} \times \big|E_n \big>
+ \omega_0^2
\big< H_m \big|\hat{f}\big|H_n \big> .
\label{Thetamn}
\end{eqnarray}
Since $\hat{\Theta}$ is a Hermitian operator we may apply the
variational principle
\begin{eqnarray}
 \omega^2\big[ H \big]
&=& \min_{\left| H \right>} \frac{\big<H \big| \hat{\Theta}
 \big| H \big>}{\big<H \big| H \big>}\nonumber\\
 &=&
\min_{\{c_n\}} \sum_{mn}c_m^* c_n \Big( -i\omega_0 \big< H_m
\big|\nabla\hat{f} \times \big|E_n \big> + \omega_0^2 \big< H_m
\big|\hat{f}\big|H_n \big> \Big) . \label{varH}
\end{eqnarray}
Thus the variational approach in $\HHH$ gives a second order error
in frequency for a first order error in the basis function. Note,
that although the $\EEE$ field appears in Eq. (\ref{varH}) this is
only a convenient way of writing the curl of an $\HHH$ field times
$\omega_0/(i\mathbf{\varepsilon}_0)$, and variational error in
$\EEE$ has no direct effect in the equation.

If first order perturbation is applied to the $\HHH$-field
eigenvalue equation, Eq.~(\ref{Theta}), we also arrive at the
result given by Eq.~(\ref{1pt}). We note that this is only
apparently contradicting Eqs.~(2) and (5) of
Ref.~\cite{johnson2002}, since these equations present the first
order perturbation results in $\Delta\varepsilon$ and
$\Delta(\frac{1}{\varepsilon})$, respectively.

%------------------------------------------------------
\section{Example: Effective index of a photonic crystal
fibre}\label{sec:example}
%------------------------------------------------------

We consider the photonic crystal fibre first studied by Knight
\emph{et al.}~\cite{knight1996} in which air-holes (of diameter
$d$) are arranged in a micron scale triangular lattice (with pitch
$\Lambda$) and the core is formed by a defect consisting of a
``missing" air hole in the otherwise periodic structure. From the
scale invariance of Maxwell equations~\cite{joannopoulos} it
follows that the dispersion properties are uniquely defined in
terms of normalized quantities, such as the normalized free-space
wavelength $\lambda/\Lambda$ and the relative air-hole diameter
$d/\Lambda$. In this application we investigate how changing the
size of the relative air-hole diameter changes the normalized
frequency, $\omega\Lambda/(c 2\pi)=\Lambda/\lambda$.

One might speculate whether such problems can be addressed with
the aid of first-order perturbation theory. However, for silica
the variations of the air-hole diameter amounts to index
variations of the order $\delta\varepsilon \simeq
\varepsilon_\textrm{Si}- \varepsilon_\textrm{Air} \simeq 2.10-1.00
= 1.10 $ so that the relative change is certainly not close to
zero as assumed in the derivation of Eq.~(\ref{1pt}). Thus this
problem certainly calls for approaches beyond ordinary
perturbation theory. Here we apply the variational approaches
suggested above.

%------------------------------------------------------
\subsection{Numerical model}
%------------------------------------------------------
In fibre optics it is common to consider the frequency dependent
effective index,
\begin{eqnarray}
n_\textrm{eff}(\omega) &\equiv&  \frac{ c\beta(\omega)}{\omega},
\label{neff}
\end{eqnarray}
rather than the usual dispersion relation $\omega(\beta)$. For
obtaining a reference basis set we employ a plane-wave method in a
super-cell configuration~\cite{johnson2001}, to provide us with
the eigenmodes at a give frequency. Using these we evaluate the
frequency shifts with  both the variational method in $\EEE$ and
$\HHH$ and diagonalize the operators to find the lowest
eigenfrequency. For the PCF the dielectric function is isotropic
so $\varepsilonbf(\rrr) = \varepsilon(\rrr)\cdot
\mathbf{1}_{3\times3}$ where $\varepsilon(\rrr)$ is a real scalar.

\begin{figure}[t]
\begin{center}
\epsfig{file=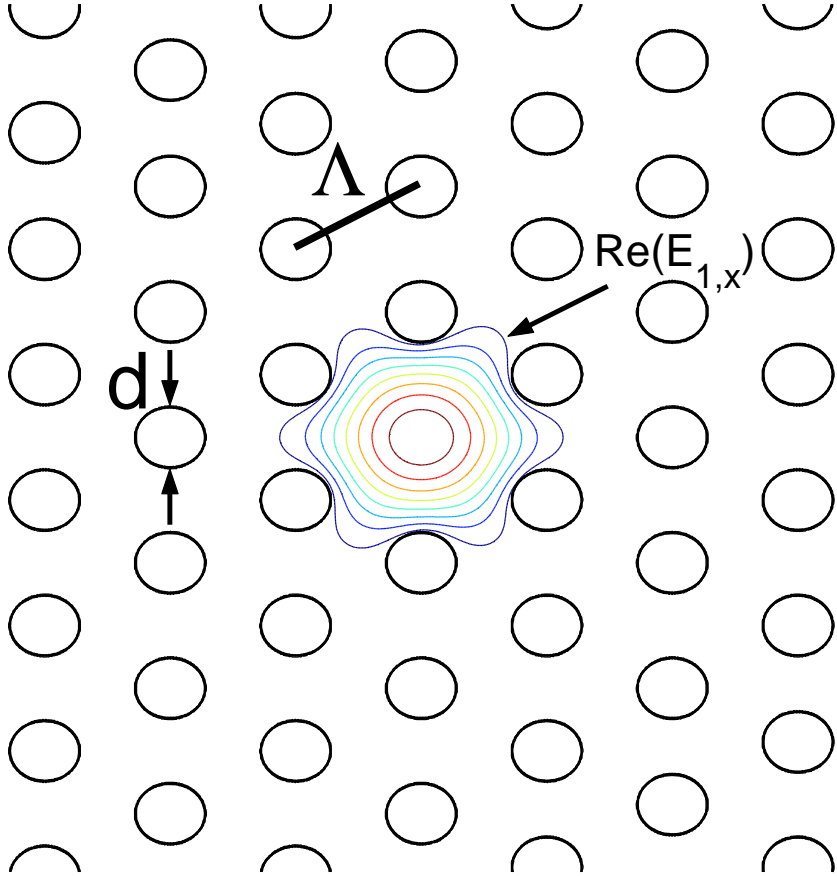, width=0.8\columnwidth,clip}
\caption{Cross section of a photonic crystal fiber with a relative
air-hole diameter $d/\Lambda=0.48$. Light can be localized to a
core region, which is formed by leaving out a single air hole in
the otherwise periodic array of air-holes. The contours show the
real part of a transverse component of the $\EEE$-field amplitude
calculated with the plane-wave method for
$\beta\Lambda=41.4$.\label{fig1} }
\end{center}
\end{figure}

A PCF has a two fold degenerate fundamental mode corresponding to
the two polarization states in free-space and cylinder-symmetrical
problems~\cite{steel2001}. In the following we will limit
ourself to a Hilbert space consisting only of these two modes
($N=2$). The PCF considered is made of a triangular structure of
holes with the relative diameter $d/\Lambda = 0.48$, see
Fig.~\ref{fig1}. We consider a normalized propagation constant of
$\beta\Lambda = 41.4$, and a refractive index for the silica of
$n_\textrm{Si}=1.45$. These structure parameters correspond to
1550 nm wavelength operation for a typical large-mode area
photonic crystal fiber \cite{Nielsen:2004} and should be
considered fairly standard for PCFs.
%------------------------------------------------------
\subsection{Numerical results}
%------------------------------------------------------
For the numerical implementation we employ a cell of $600\times
600\times 1 = 36000$ grid points. The minimal length between the
cores in the super-cell images is $6\Lambda$. For comparison we do
a full calculation for each hole size. In these calculations we
set the propagation constant $\beta$ in Eq.~(\ref{neff}) to a
fixed value and calculate the corresponding frequency.

\begin{figure}[t]
\begin{center}
\epsfig{file=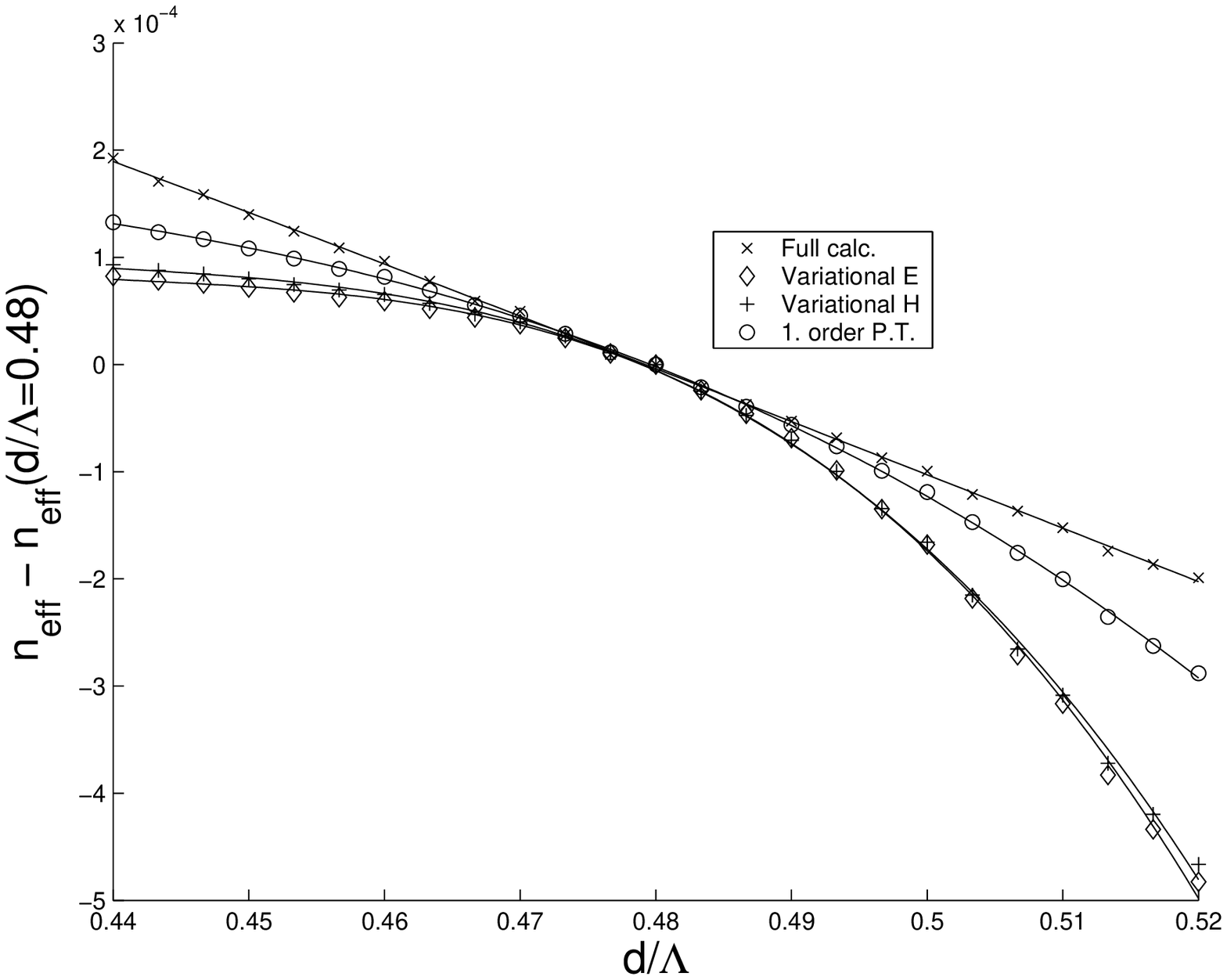,width=\columnwidth,clip}
\caption{Effective index versus normalized air-hole diameter of
the fundamental mode in a photonic crystal fiber calculated for
$\beta\Lambda= 41.4$. The upper curve shows results of a
plane-wave simulation while the three other curves show the
presented variational approaches taking the plane-wave simulation
for $d/\Lambda=0.48$ as a starting point. } \label{fig2}
\end{center}
\end{figure}

The calculation time for the full calculation with the plane-wave
code is 25$\times$10 minutes (6 hours) on a dedicated 1.4 GHZ
machine. The calculation time for the presented method is 10
minutes for a single full calculation, 3 minutes for outputting
the dielectric functions with the same program, and 1 minute for
calculating the frequency shifts. The program was run on a desktop
866 MHZ machine and file I/O consumed the larger part of the time.
The dielectric functions used in the presented method were the
same as were used in the full calculation, to give maximal
consistency between the results of the variational method and the
full calculations.

In Fig.~\ref{fig2} the calculated effective index versus
normalized air-hole diameter is presented. It is seen that the
error is of second order in the hole size. This is due to the
variational principle Eqs.~(\ref{varE},\ref{varH}) which causes
the frequency $\omega$ to be overestimated leading to an
$n_\textrm{eff}\sim\omega^{-1}$ that is too small.

%------------------------------------------------------
\subsection{Discussion}
%------------------------------------------------------
The cause of the error in our variational methods is the
discrepancy between the exact eigenmodes and the eigenmodes of the
reference system.

The electrical and magnetic fields are rapidly decaying inside the
air holes with an exponential tail. So when the air hole boundary
is moved a considerable distance, the reference fields deviate
considerably from the exact fields. The condition for an electric
charge free system, $\nabla\cdot[ \varepsilon(\rrr)\EEE(\rrr)] =
0$, is not fulfilled for the `new' system in the $\EEE$
variational method, whereas the $\HHH$ field is not subjected to
the same condition in the $\HHH$ variational method, since we have
$\nabla\cdot[ \mu \HHH(\rrr)] = 0$ . This is probably the cause to
the $\HHH$ variational method being slightly closer to the full
calculation in Fig.~\ref{fig2}. The reference fields in both
variational methods thus do not describe the actual physical
problem, but since we are interested only in eigenfrequencies and
not their corresponding fields, this does not pose a problem since
the error in eigenfrequencies is of second order in the field.

One might speculate why first order perturbation theory seems to
perform better than the variational method as indicated by Fig.
(\ref{fig2}). This can be explained by observing
Eqs.~(\ref{1ptexpansion}) and (\ref{1pt}). In our problem the
numerically evaluation of the matrix element in the first line
Eq.~(\ref{1ptexpansion}) is an overlap integral (over the area
given by $\varepsilon(\rrr) - \varepsilon_0(\rrr)\neq0$) times a
prefactor which is proportional to
$\varepsilon_0(\rrr)/(\varepsilon_0(\rrr)+\delta\varepsilon(\rrr)-1$
which is constant in the $\varepsilon(\rrr) -
\varepsilon_0(\rrr)\neq0$ area. . Since integrals for first order
perturbation theory and the variational approaches evaluate to the
same value, but the prefactor is $-\delta\varepsilon(\rrr)$ in the
second line of Eq.~(\ref{1ptexpansion}), leading to first order
perturbation theory to spuriously correct the variational error,
by over- and underestimating the frequency for $d/\Lambda<0.48$
and $d/\Lambda>0.48$, respectively. Theoretically, first order
perturbation theory is not correct to any order for high-index
contrast systems.

One could include a couple of extra eigenmodes for an improved
basis, but beyond the two fundamental modes, the following modes
poorly describe changes due to varying hole size. We could have
included all $600\times 600\times 1 = 360000$ eigenmodes in the
variational calculation, but this would be equivalent to a full
calculation, and thus would make only little sense. One could,
though, include eigenmodes that are confined around the hole
boundaries for an improved description. This would give a much
better estimation of the frequency changes.

Finally, it is also possible to reduce the calculation time of the
reference system by using a coarser grid, and then afterwards
interpolate the fields to a higher resolution.

%------------------------------------------------------
\section{Conclusion}\label{sec:conclusion}
%------------------------------------------------------

We have presented a method that can calculate frequency shifts for
arbitrary changes in the refractive index, both isotropic and
anisotropic, without complications due to degeneracy of the
eigenfrequencies. We have demonstrated that the method can
estimate frequency shifts which are exact to first order in the
basis set.

The method is capable of calculating frequency shifts
which are too small to be calculate efficiently with a full
calculation. The theory of the method is transparent and provides
a sound basis for more elaborate numerical models.

The list of potential applications of the presented theory
includes fiber gratings, birefringence of the doubly degenerate
fundamental mode in PCFs due to structural
inhomogeneities~\cite{mortensen2004b}, and speed up for structural
topological optimizations of electromagnetic
systems~\cite{Jensen:2004b}.

\vspace{1cm}
\section*{Acknowledgement}
We thank Steven G. Johnson and Michael Pedersen for discussions
and useful comments on the initial manuscript. N.~A.~M. is
supported by The Danish Technical Research Council (Grant
No.~26-03-0073).

%\bibliographystyle{elsart-num}
%\bibliography{pcf}

\end{document}